\documentclass[9pt,twocolumn,twoside]{osajnl}

\journal{preprint} 

\setboolean{shortarticle}{true}

\title{Fabrication-Robust Silicon Photonic Devices in Standard Sub-Micron Silicon-on-Insulator Processes}

\author[1,*]{Anthony~Rizzo}
\author[1]{Utsav~Dave}
\author[1]{Asher~Novick}
\author[1]{Alexandre~Freitas}
\author[1]{Samantha~P.~Roberts}
\author[1]{Aneek~James}
\author[1]{Michal~Lipson}
\author[1]{Keren~Bergman}

\affil[1]{Department of Electrical Engineering, Columbia University, New York, NY, 10027}

\affil[*]{Corresponding author: anthony.rizzo@columbia.edu}

\begin{abstract}
Perturbations to the effective refractive index from nanometer-scale fabrication variations in waveguide geometry plague high index-contrast photonic platforms including the ubiquitous sub-micron silicon-on-insulator (SOI) process. Such variations are particularly troublesome for phase-sensitive devices such as interferometers and resonators, which exhibit drastic changes in performance as a result of these fabrication-induced phase errors. In this Letter, we propose and experimentally demonstrate a design methodology for dramatically reducing device sensitivity to silicon width variations. We apply this methodology to a highly phase-sensitive device, the ring-assisted Mach Zehnder interferometer (RAMZI), and show comparable performance and footprint to state-of-the-art devices while substantially reducing stochastic phase errors from etch variations. This decrease in sensitivity is directly realized as energy savings by significantly lowering the required corrective thermal tuning power, providing a promising path towards ultra-energy-efficient large-scale silicon photonic circuits. 
\end{abstract}

\setboolean{displaycopyright}{true}

\begin{document}

\maketitle


Silicon photonics has recently emerged as a leading technology beyond its initial intended market of optical interconnects to additionally include diverse applications such as LIDAR, deep learning accelerators, medical sensors, and quantum computers. To address all of these markets with a standardized, general-purpose process, the majority of leading foundries have converged on sub-micron SOI platforms with typical silicon device layer heights ranging from 220 nm to 310 nm \cite{AMF, AIMPDK}. To maintain single mode operation with these heights, typical waveguide widths range from $\approx 300$ nm for O-band ($\lambda = 1310$ nm) to $\approx 450$ nm for C-band ($\lambda = 1550$ nm). All of the aforementioned applications rely on the standard integrated photonics toolbox of devices, of which the main workhorses are resonators and interferometers. However, in sub-micron SOI processes, the performance of these devices is highly susceptible to minute changes in the waveguide width and height on the order of a few nanometers.

While sensitivity to silicon thickness variations cannot be mitigated since the waveguide height dimension is fixed by the process, the waveguide width is lithographically defined and is thus a degree of freedom. It is well known that wider waveguides are less susceptible to variations in width \cite{Song:21}; however, this is at the expense of supporting undesirable higher-order modes and thus it was previously thought that these wide waveguides could only be used to reduce phase errors in long, straight sections of devices \cite{Song:21, mzi_ng}. However, it is possible to maintain pure single mode operation for wide multi-mode waveguide bends in sub-micron processes by adiabatically varying the radius of curvature to ensure that the mode discontinuities are minimized and thus higher order modes remain unexcited. Since wide waveguides have the additional benefit of reduced propagation losses due to less overlap between the optical mode and rough sidewalls, previous work with wide multi-mode waveguides operating in the single mode regime with adiabatic bends has focused on enabling ultra-high quality factor (Q) resonators rather than fabrication-robust devices \cite{Lipson2021, HighQ, kippenberg}. Previous work in thick silicon processes (> 1 $\mu$m) has demonstrated the use of adiabatic curves to reduce the excitation of higher order modes in bends \cite{cherchi2013dramatic} and in general thick silicon processes are more fabrication-robust than sub-micron processes, but they are far less commonly used and are restricted to a few highly specialized foundries \cite{Rockley, VTT}. 

In this work, we employ Euler curves (also commonly referred to as clothoid curves \cite{euler}) with wide multi-mode waveguides in a standard 220 nm silicon photonics platform to show that complex single mode devices with compact bends and resonators can be constructed using fabrication-robust wide waveguides without degradation in performance or increase in footprint. We demonstrate fabrication-robust dense wavelength-division multiplexing (DWDM) RAMZI interleavers as a representative proof-of-principle device with state-of-the-art performance and footprint. Furthermore, we provide a comprehensive design space exploration using finite difference eigenmode (FDE) and rigorous 3D finite difference time domain (FDTD) simulations to examine the trade-offs in Euler bend designs for different widths and identify target design points. Finally, we fabricated RAMZI devices using electron beam (e-beam) lithography in a university clean room setting as well as deep ultraviolet (DUV) lithography in a commercial 300 mm foundry, demonstrating the universality of the methodology and its natural compatibility with high-volume fabrication. This demonstrated general design methodology illuminates an appealing path towards large-scale silicon photonic circuits that require substantially less thermal tuning power to correct for stochastic phase errors when compared to conventional designs.


\begin{figure}[t!]
    \centering
    \includegraphics{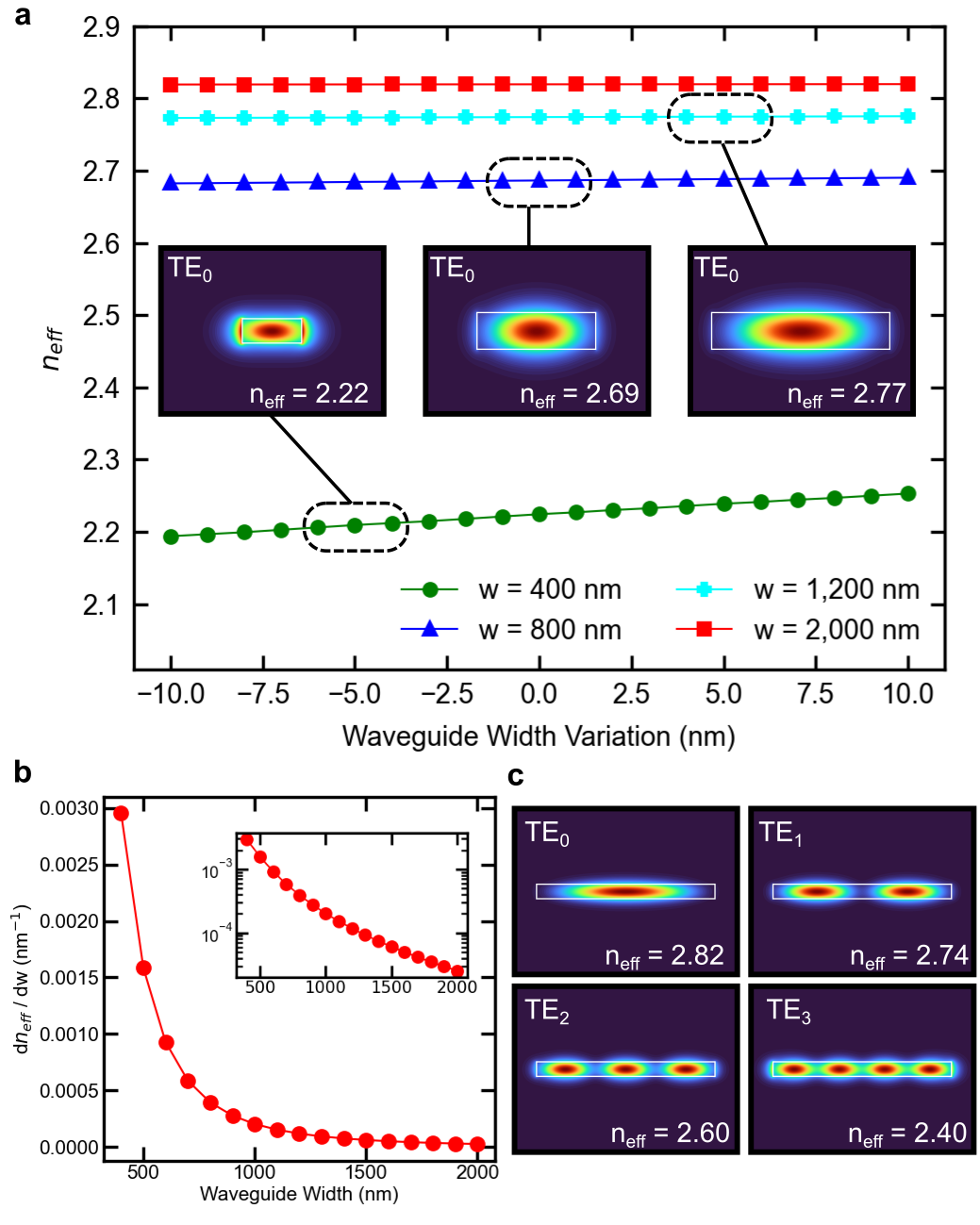}
    \caption{\textbf{a,} Simulated effective refractive index (n\textsubscript{eff}) of the fundamental transverse electric (TE) mode as a function of width variation for various nominal widths. Insets: Corresponding simulated fundamental mode profiles for the circled widths. \textbf{b,} Sensitivity of n\textsubscript{eff} to width variations as a function of nominal width. Inset: Identical data plotted on a logarithmic scale. \textbf{c,} Simulated mode profiles for the first four supported modes of a $2,000 \times 220$ nm SOI waveguide. Note that all simulations are for $\lambda = 1,550$ nm and different wavelengths will experience marginally different sensitivities following the same trend line.}
    \label{fig:width_variations}
\end{figure}

For sub-micron SOI waveguides with a fixed height of 220 nm, we first use FDE simulations to explore the relationship between nominal waveguide width and sensitivity to width variations. Fig. \ref{fig:width_variations}a shows the sensitivity in effective refractive index (n\textsubscript{eff}) of various waveguide geometries to variations in width. The simulated width variations of $\pm 10$ nm are well within the $3\sigma$ values for wafer-scale measured data from dedicated silicon photonics foundries \cite{AMF, AIMPDK}. Since the curves from Fig. \ref{fig:width_variations}a are linear, the sensitivities $\partial$n\textsubscript{eff} / $\partial w$ are constant and are plotted in Fig. \ref{fig:width_variations}b as a function of nominal width. From these simulations, it is clear that the widely used conventional single mode waveguides ($w = 400$ to $500$ nm) are highly sensitive to width variations, with a sensitivity of 3 $\times 10^{-3}$ nm\textsuperscript{-1} at $w = 400$ nm ($\lambda = 1,550$ nm). However, we can also see that through using wider waveguides, this sensitivity can be dramatically reduced by over two orders of magnitude to 2.5 $\times 10^{-5}$ nm\textsuperscript{-1} for $w = 2,000$ nm.

This advantage comes with a caveat, as wide waveguides begin to support a plethora of higher-order spatial modes. The first four transverse electric (TE) modes of a $2,000 \times 220$ nm SOI waveguide are shown in Fig. \ref{fig:width_variations}c with their corresponding effective indices. In typical devices and circuits, it is highly undesirable to excite these modes through parasitic conversion of light from the fundamental mode as it results in increased losses and degraded performance. Since the dominant source of this parasitic conversion is in waveguide bends, it is standard to only use wide waveguides in long, straight sections with tapers on both ends to interface with the single mode waveguides used in the rest of the circuit. However, as mentioned previously, through carefully designing the bends for adiabatic mode propagation the entire circuit can employ wide waveguides without degrading performance and without the need for tapers. An additional important consideration when choosing width for broadband applications is the total dispersion experienced by the target guided mode, which is influenced both by material dispersion and waveguide dispersion. For oxide clad silicon waveguides with 220 nm height, we find from FDE simulations that minimum dispersion occurs around 640 nm width and worsens monotonically as the width is increased further. This presents an application-dependent design choice, as Figure \ref{fig:width_variations}b shows that wider waveguides beyond this point are more robust to fabrication variations, but this comes at the expense of experiencing higher dispersion. 

\begin{figure}[ht!]
    \centering
    \includegraphics{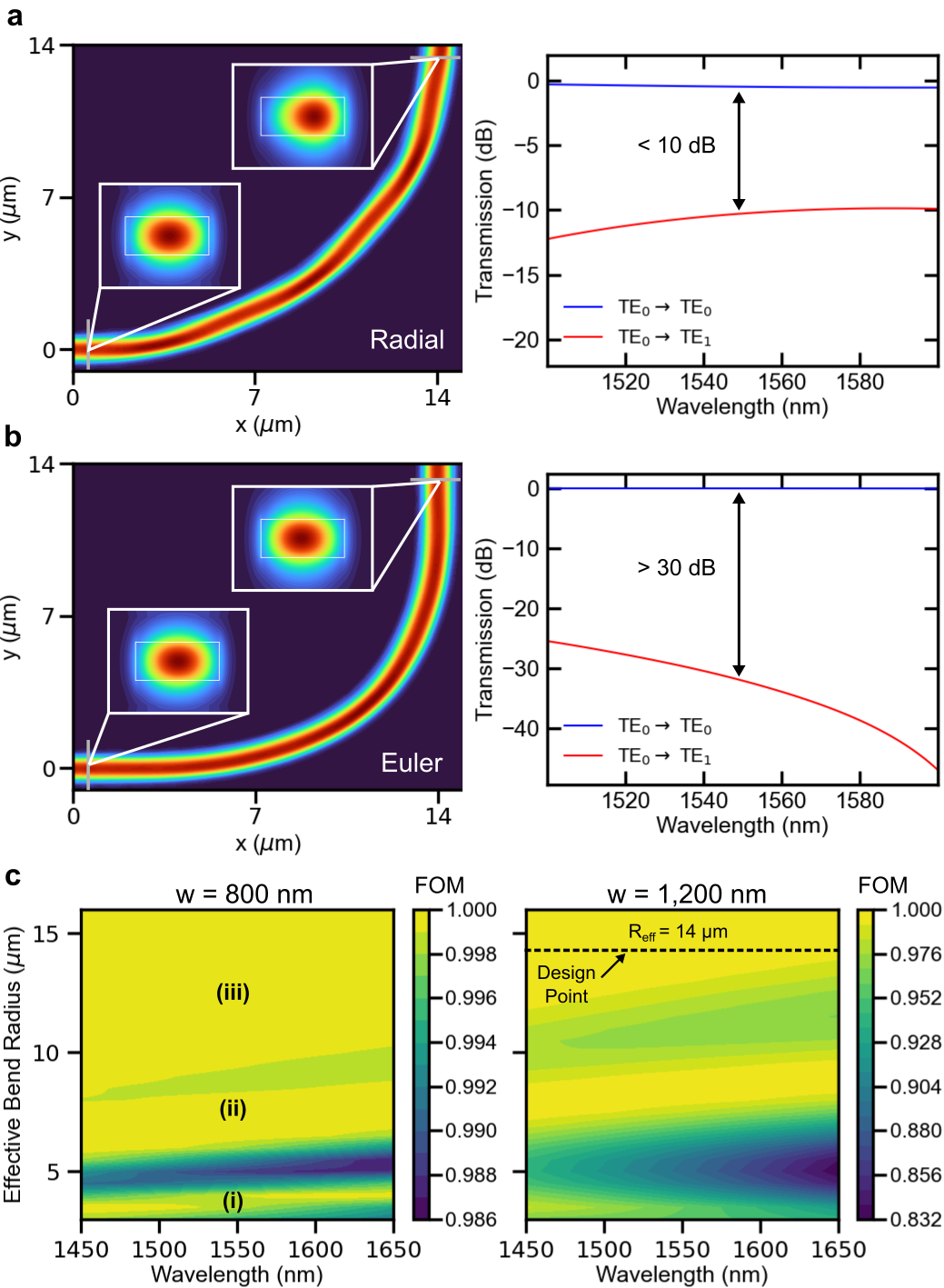}
    \caption{\textbf{a,} Simulated field profile for a radial bend with R = 14 $\mu$m and $w = 1,200$ nm (Insets: mode profiles at the input and output) and its corresponding scattering parameters for transmission into the first two supported modes. \textbf{b,} Simulated field profile for an Euler bend with R\textsubscript{eff} = 14 $\mu$m ($w = 1,200$ nm) and its corresponding scattering parameters. \textbf{c,} Simulated TE\textsubscript{0}$\rightarrow$TE\textsubscript{0} transmission as a function of R\textsubscript{eff} for 800 nm and 1,200 nm wide waveguides.}
    \label{fig:euler}
\end{figure}

In Cartesian coordinates, a radial bend is parameterized by the relation $x^{2} + y^{2} = R^{2}$, where $R$ is a constant radius of curvature. In contrast, Euler bends have a radius of curvature that varies linearly along the path length, defined by the Fresnel integrals \cite{nitride_euler, mb}
\begin{equation}
    x(s) = \int_{0}^{s} \cos\left(\frac{t^{2}}{2 R_{0}^{2}}\right) dt
\end{equation}
\begin{equation}
    y(s) = \int_{0}^{s} \sin\left(\frac{t^{2}}{2 R_{0}^{2}}\right) dt
\end{equation}
where $s$ is the normalized path length, $R_{0}$ is the parameter of the Euler curve, and the curvature function is $\kappa(s) = \frac{2s}{R_{0}^2}$. Restricting our analysis to 90 degree bends, which consist of 2 concatenated 45 degree Euler spirals, it is helpful to write $R_{0}$ in terms of the minimum bend radius $R_{min}$ as \cite{mb} $R_{0} = \sqrt{2 R_{min} s_{mid}}$, where $s_{mid}$ is the half length of the full curve. Since the radius of curvature changes linearly along the path, abrupt modal discontinuities are minimized and the transition through the bend is adiabatic for an appropriate choice of $R_{0}$.  

Using 3D FDTD simulations, we explore the design space of bends for various widths and compare the performance of Euler bends versus radial bends (Fig. \ref{fig:euler}). The figure of merit (FOM) for these bends is TE\textsubscript{0}$\rightarrow$TE\textsubscript{0} transmission, with sub-unity values indicating losses due to mode-mismatch, bend radiation, and mode conversion. To use total footprint as a basis of comparison, we define the effective bend radius R\textsubscript{eff} of an Euler bend to be an Euler bend with the same (x, y) dimensions as a radial bend of radius R = R\textsubscript{eff}. From these simulations, we identify `safe' regions of operation for Euler bends at each nominal width and observe three distinct regimes for TE\textsubscript{0}$\rightarrow$TE\textsubscript{0} transmission as a function of R\textsubscript{eff} (Fig. \ref{fig:euler}c). Interestingly, regime (i) shows good figure-of-merit performance for extremely small Euler bends below R\textsubscript{eff} = 4 $\mu$m, which then worsens as R\textsubscript{eff} is increased. Further increasing R\textsubscript{eff} beyond this region improves the FOM in regime (ii), before degrading again and finally showing monotonically improving performance in regime (iii). From the observed field profiles, the performance in regimes (i) and (ii) appears to benefit from multi mode interference rather than adiabatic mode propagation and thus we focus our designs on regime (iii). Future work is necessary to experimentally explore the efficacy of using regimes (i) and (ii) to further reduce the bend footprint.


We choose the RAMZI as a representative device for our methodology since it is of current interest for use in DWDM link architectures \cite{rizzo2021integrated, HPE}, highly phase-sensitive, contains both resonant and delay-imbalanced interferometric elements, and requires a large power transfer between adjacent waveguides in a compact footprint. The final point presents a nuanced but substantial challenge, as the increased confinement of wide waveguides results in a much smaller evanescent field and precludes the use of evanescent directional couplers for compact structures (previous demonstrations have required couplers with lengths of hundreds of microns to millimeters for $\approx$ 1\% to 5\% power transfer \cite{Lipson2021, HighQ}). For a single-ring-loaded RAMZI interleaver, the power coupling coefficient for optimal passband flatness and crosstalk suppression is $\kappa = 0.89$ \cite{Rizzo_PTL_2021, HPE}, which entirely eliminates the possibility of using directional couplers while maintaining a compact device footprint. Instead, we use multi-mode interference couplers (MMIs) designed with arbitrary splitting ratios to transfer light between waveguides, including from the Mach Zehnder arm to the ring \cite{VTT, ring_mmi, Rizzo_GFP}. MMIs provide a natural splitting/combining element in our wide waveguide platform since standard designs typically taper from $\approx$ 400 nm to $\approx$ 1.2 $\mu$m before injecting light into the multi-mode body region, which provides lower loss, increased bandwidth, and fabrication-robust performance \cite{mmi}. In our platform, this allows us to directly abut the input/output waveguides to the MMI body and entirely eliminate tapers.


\begin{figure}
    \centering
    \includegraphics{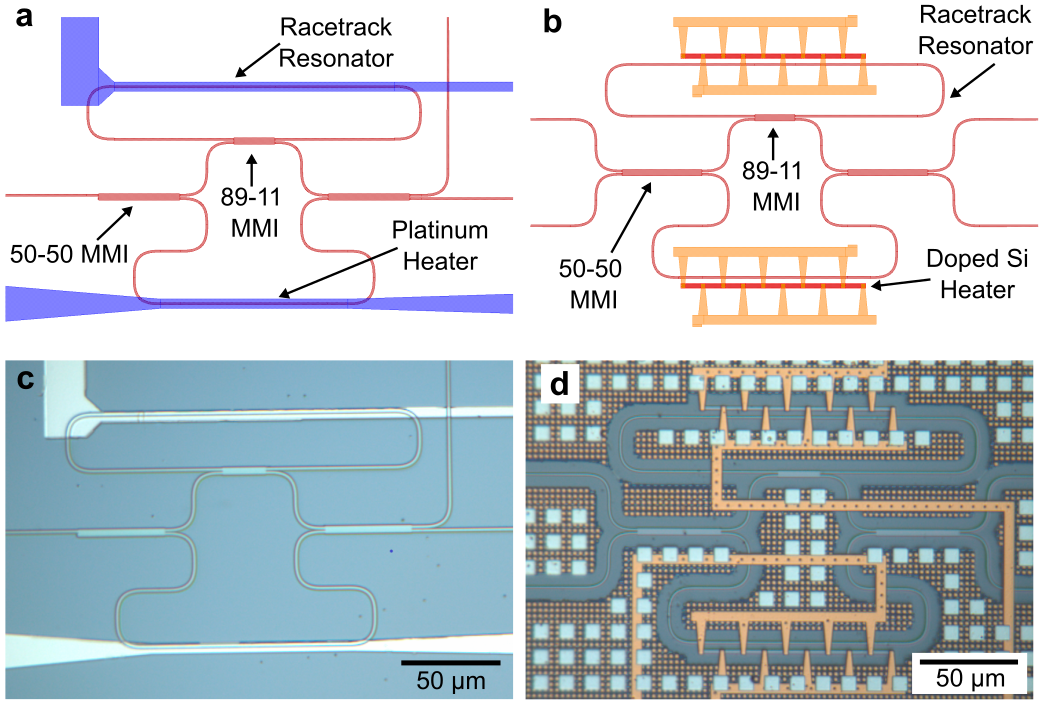}
    \caption{\textbf{a,} Annotated mask layout of the e-beam device. \textbf{b,} Layout of the foundry-fabricated device. \textbf{c,} Optical microscope image of the e-beam device. \textbf{d,} Optical microscope image of the foundry-fabricated device.}
    \label{fig:gds}
\end{figure}

\begin{figure}
    \centering
    \includegraphics{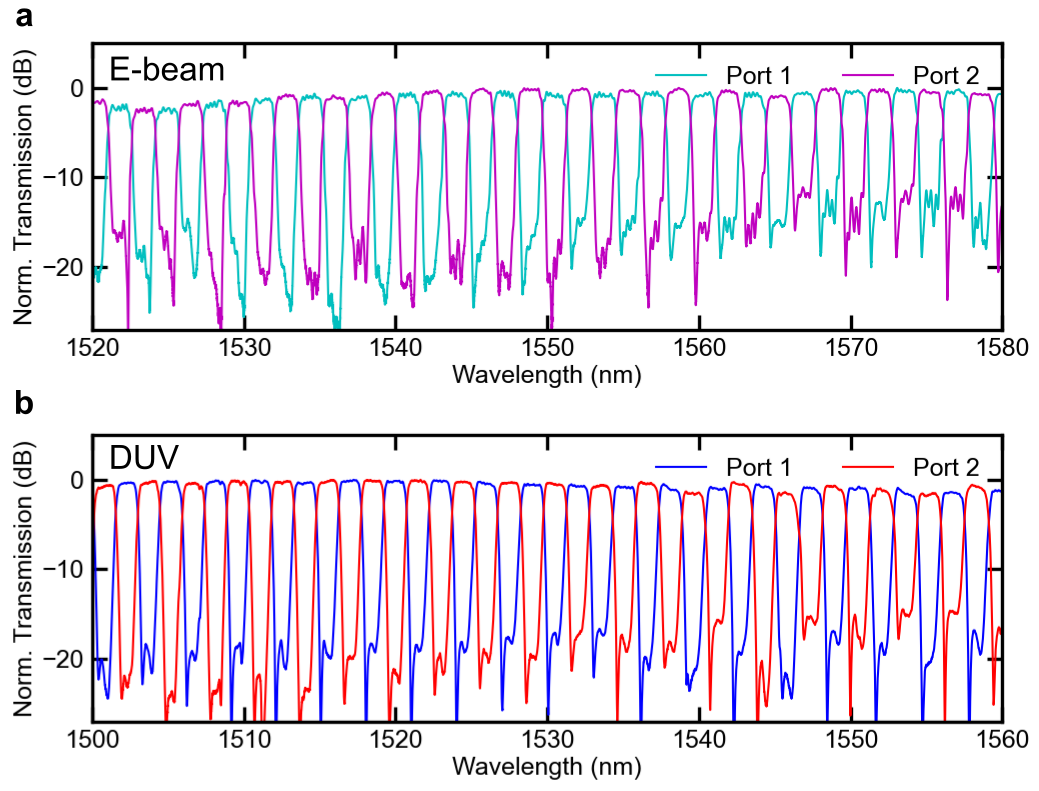}
    \caption{\textbf{a,} Experimentally measured spectrum for the e-beam device, displaying flat top pass- and stop-bands with typical crosstalk of -20 dB and worst-case crosstalk of -12 dB. \textbf{b,} Measured spectrum for the foundry-fabricated device, with comparable pass- and stop-band flatness and crosstalk suppression ratios. Note that for the foundry-fabricated device, the ideal transmission band is blue-shifted $\approx$ 20 nm which is likely due to etch bias in the MMI body and requires re-zeroing.}
    \label{fig:spectra}
\end{figure}

\begin{figure}
    \centering
    \includegraphics{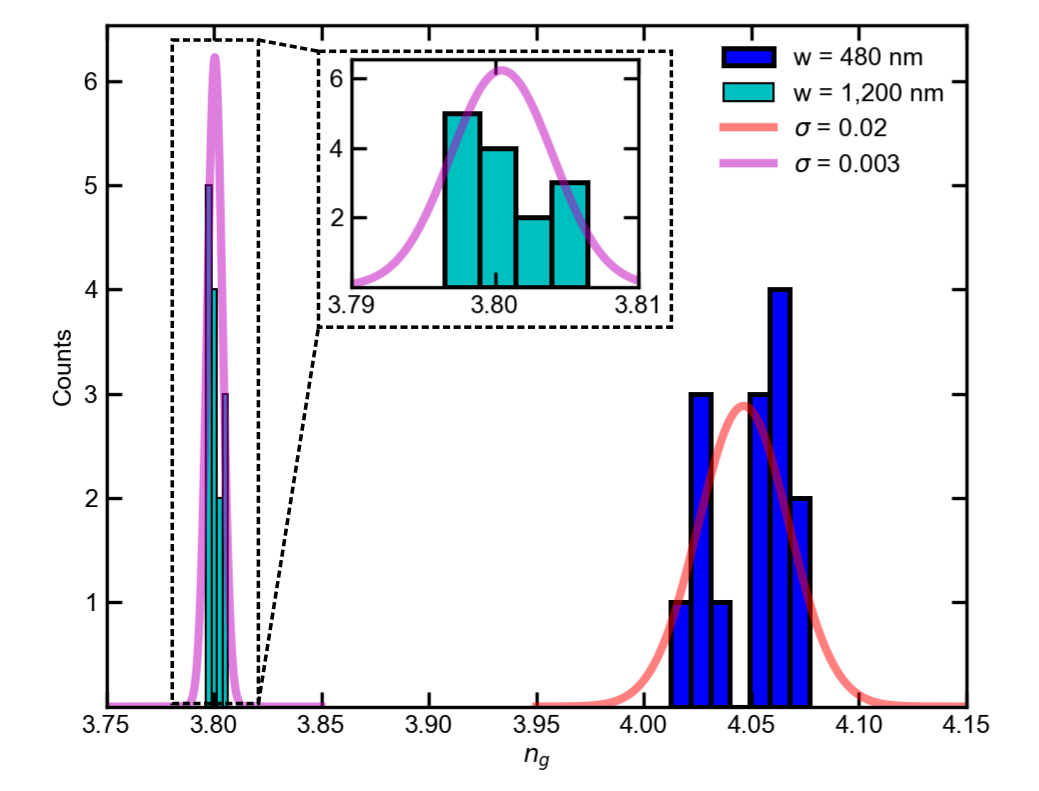}
    \caption{Measured statistical variation in n\textsubscript{g} across 14 standard width and 14 wide devices. The wide devices show nearly an order of magnitude reduction in standard deviation ($\sigma$).}
    \label{fig:statistics}
\end{figure}

For our devices, we chose a nominal waveguide width of 1,200 nm to obtain an optimal balance between robustness to fabrication variations and dispersion for broadband applications. Both MMIs maintain this width for all ports with center-to-center waveguide spacing of 1.8 $\mu$m; the 89-11 MMI has body dimensions of 3.5 $\mu$m $\times$ 21.6 $\mu$m and the 50-50 MMI has body dimensions of 3.5 $\mu$m $\times$ 43.1 $\mu$m. The footprint of the entire structure is only 0.02 mm\textsuperscript{2} (Fig. \ref{fig:gds}). For the e-beam devices, we implemented thermo-optic phase shifters as a platinum layer above the waveguide. Due to the absence of a metal heater layer in the DUV foundry process used, we implemented the heaters for these devices as a doped silicon resistor adjacent to the waveguide. Both the foundry-fabricated and e-beam devices display performance comparable to the state-of-the-art in terms of footprint, crosstalk, bandwidth, and pass-band shape \cite{HPE} (Fig. \ref{fig:spectra}). We measured a device temperature sensitivity of 9.74 GHz/K, which closely matches the simulated value of 8 GHz/K. The foundry-fabricated devices were taped out as part of the AIM Photonics 300 mm multi-project wafer (MPW) run \cite{AIMPDK}. Since the die were singulated prior to shipping, the exact location of each die on the wafer was unknown and thus we were unable to control for height variations across devices. Thus, the statistics shown in Fig. \ref{fig:statistics} are agnostic to die location and therefore include the effects of both height and width variations. Nevertheless, it is still clear that the wide waveguide devices display substantially less stochastic phase errors than the nominal width designs. Since a full 2$\pi$ phase error (which we assume is possible within the bounds of fabrication variations) will shift the entire spectrum a full FSR and alias with the non-shifted spectrum, we opt to calculate the group index n\textsubscript{g} from the FSR to quantify fabrication variations rather than make assumptions about the fringe order to extract n\textsubscript{eff} \cite{mzi_ng}. Planned future work will include a dedicated 300 mm wafer with comprehensive test structures for location-aware wafer-scale testing to further quantify statistical distributions as a function of waveguide width. 


In summary, we have demonstrated a design methodology which greatly reduces stochastic phase errors in phase-sensitive silicon photonic devices without any process changes. Through mitigating these phase errors at the design stage, this approach is fully passive and inherently high-throughput as it does not require any post-processing steps such as trimming. Furthermore, the platform can naturally be integrated with ultra-high efficiency phase shifters such as thermal undercut heaters \cite{undercut} and heterogeneous III-V/Si metal-oxide-semiconductor capacitor (MOSCAP) structures \cite{HPE}. In the case of undercut heaters, our analysis shows that an average required tuning power of $< 100$ $\mu$W per device can be expected when combined with our demonstrated platform and only requires the addition of a single mask to the fabrication process. We envision that these results will influence the design of silicon photonic circuits across a broad application space, resulting in large-scale systems with dramatically reduced energy consumption compared to previous standards.

\begin{backmatter}
\bmsection{Funding} ARPA-E ENLITENED (DE-AR000843), DARPA PIPES (HR00111920014).

\bmsection{Acknowledgments} The authors thank AIM Photonics for fabrication, Analog Photonics for PDK support, and Kaylx Jang for chip dicing.

\bmsection{Disclosures} The authors declare no conflicts of interest.

\bmsection{Data availability} Data underlying the results presented in this paper are not publicly available at this time but may be obtained from the authors upon reasonable request.

\end{backmatter}

\bibliography{main}

\bibliographyfullrefs{main}


\end{document}